\documentclass{aastex}
\usepackage{emulateapj5}
\begin{document}

\newcommand{\hi}{$h^{-1}$~}
\newcommand{\kms}{~km~s$^{-1}$}
\newcommand{\logh}{+5log$h$}

\title{Field E+A Galaxies at Intermediate Redshifts $(0.3<z<1)^{1,2}$}

\author{Kim-Vy H. Tran$^3$, Marijn Franx$^4$, Garth D. Illingworth$^5$,\\
Pieter van Dokkum$^6$, Daniel D. Kelson$^7$, \& Daniel Magee$^5$ }

\altaffiltext{1}{Based on observations with the NASA/ESA Hubble Space
  Telescope, obtained at the Space Telescope Science Institute, which
  is operated by the Association of Universities for Research in
  Astronomy, Inc., under NASA contract NAS 5-26555.}
\altaffiltext{2}{Based on observations obtained at the W. M. Keck
  Observatory, which is operated jointly by the California Institute of
  Technology and the University of California.}

\altaffiltext{3}{Institute for Astronomy, ETH H\"onggerberg, CH-8093
  Z\"urich, Switzerland, vy@phys.ethz.ch}
\altaffiltext{4}{Leiden Observatory, P.O. Box 9513, 2300 RA Leiden,
    The Netherlands}
\altaffiltext{5}{University of California Observatories/Lick
    Observatory, University of California, Santa Cruz, CA 95064}
\altaffiltext{6}{Department of Astronomy, Yale University, New Haven,
  CT 06520-8101}
\altaffiltext{7}{Observatories of the Carnegie Institution of
  Washington, 813 Santa Barbara Street, Pasadena, CA, 91101}

\begin{abstract}
  
  We select E+A candidates from a spectroscopic dataset of $\sim800$
  field galaxies and measure the E+A fraction at $0.3<z<1$ to be
  $2.7\pm1.1$\%, a value lower than that in galaxy clusters at
  comparable redshifts ($11\pm3$\%).  HST/WFPC2 imaging for five of
  our six E+A's shows they have a heterogeneous parent population:
  these E+A's span a range in half-light radius ($0.8<r_{1/2}<8$\hi
  kpc) and estimated internal velocity dispersion
  ($50\lesssim\sigma_{est}\lesssim220$\kms), and they include luminous
  systems ($-21.6\leq M_{Bz}-5logh\leq -19.2$).  Despite their
  diversity in some aspects, the E+A's share several common
  characteristics that indicate the E+A phase is an important link in
  the evolution of star-forming galaxies into passive systems: the
  E+A's are uniformly redder than the blue, star-forming galaxies that
  make up the majority of the field, they are more likely to be
  bulge-dominated than the average field galaxy, and they tend to be
  morphologically irregular.  We find E+A's make up $\sim9$\% of the
  absorption line systems in this redshift range, and estimate that
  $\gtrsim25$\% of passive galaxies in the local field had an E+A
  phase at $z\lesssim1$.

\end{abstract}

\section{Introduction} 

Studies show the universal star formation rate has been declining
steadily since $z\sim1$ \citep{lilly:95,cowie:96,madau:96}.  However,
{\it why} star formation in the universe has been decreasing for the
last $\sim8$ Gyr and why it ends for any given galaxy remain unanswered
questions.  Galaxies whose star formation ends abruptly in a burst can
provide insight into these issues because they may be the link between
star-forming and passive systems.

Identified in significant numbers in intermediate redshift galaxy
clusters \citep[hereafter T03a]{dressler:83,couch:87,tran:03b}, ``E+A''
galaxies are characterized by strong Balmer absorption lines and weak
or absent [OII]$\lambda3727$ emission.  They have been studied
extensively in clusters
\citep[T03a]{dressler:83,couch:87,franx:93,wirth:94,couch:94,belloni:95,couch:98,fisher:98,caldwell:99,dressler:99,balogh:99,poggianti:04},
and are also referred to as H$\delta$-strong or ``k+a'' galaxies.
Their distinctive spectral features are usually interpreted as being
the result of a starburst ($M_{SB}\gtrsim0.1M_{\ast}$, where $M_{\ast}$
is the galaxy's final total stellar mass) that ended within the last
$1.5-2$ Gyr \citep[although see Newberry et al. 1990 for an alternative
scenario]{couch:87,barger:96}.

In addition to being in rich clusters, E+A galaxies have also been
found in the field.  The properties of field E+A's have been studied
extensively at low redshift
\citep[$z<0.2$;][]{liu:95,zabludoff:96,chang:01,norton:01,goto:03,quintero:04,yang:04},
but these low redshift field E+A's do not trace star formation beyond
$z\sim0.3$ because of the short lifetime of the E+A phase.  To
determine how important a role the E+A phase has in the evolution of
field star-forming galaxies into passive systems, a population of field
E+A galaxies at $z>0.3$ is needed.

Finding E+A galaxies is challenging because the E+A phase is visible
for only $\sim1.5$ Gyr \citep{couch:87,barger:96} and E+A's are only a
small fraction of the total galaxy population at any given redshift.
These difficulties are reflected in the highly uncertain estimates for
the field E+A fraction at $z>0.3$ where values range from $\sim1-5$\%
\citep{hammer:97,dressler:99,balogh:99}.  Estimates of the field E+A
fraction also can vary due to, e.g.  selection criteria and spectral
resolution.  Assuming the E+A phase is associated with the
transformation of an emission line galaxy into an absorption line
system, a high E+A fraction would imply that the majority of passive
galaxies in the nearby universe go through an E+A phase.

Given the formidable obstacle of even isolating the field E+A
population at $z>0.3$, it is no surprise then that we know very little
about their physical characteristics and their progenitors.  Studies of
E+A's in the nearby universe
\citep{zabludoff:96,caldwell:99,galaz:00,norton:01,poggianti:04,yang:04}
and in intermediate redshift clusters
\citep[T03a]{wirth:94,dressler:99} indicate these E+A's have a
heterogeneous parent population.  By examining their physical
properties, we can determine if E+A's in the intermediate redshift
field are descended from a diverse population as well.  This
information also enables us to determine whether field E+A's at $z>0.3$
can be the progenitors of early-type galaxies at lower redshifts.

What triggers the E+A phase remains an open question.  Earlier studies
suggest the E+A phase in the field is associated with galaxy-galaxy
interactions \citep[hereafter Z96]{liu:95,yang:04,zabludoff:96}, a
scenario consistent with galaxy formation models that predict
interactions in the field are common \citep{somerville:99a}, and that
interactions trigger strong star formation episodes in gas-rich systems
\citep{mihos:94a,mihos:94b}.  Because morphological signatures due to
interactions remain visible for the duration of the E+A phase
\citep{mihos:94a,mihos:94b}, we should find E+A's to be morphologically
disturbed if the E+A phase is associated with galaxy-galaxy
interactions.

To measure the field E+A fraction at intermediate redshifts, we draw
from an extensive spectroscopic survey of four different fields.  Our
unique dataset contains an unsually large number of spectroscopically
confirmed field galaxies ($\sim800$) gathered over a total area of
$\sim200\Box'$; our survey is larger than that of any other previous
field E+A study in this redshift range.  Using the wide-field HST/WFPC2
imaging that we have obtained as part of our cluster program, we also
characterize the physical properties of the field E+A population to
better understand the role E+A's have in the evolution of star-forming
galaxies into passive systems.

We summarize our observations in \S2, describe our E+A selection
criteria and determine the field E+A fraction in \S3, and detail the
physical properties of the field E+A population in \S4.  We discuss
the importance of the E+A phase in \S5 and present our conclusions in
\S6. Unless otherwise noted, we use $\Omega_M=0.3,
\Omega_{\Lambda}=0.7$, and $H_0=100h$\kms~Mpc$^{-1}$ in this paper.

\section{Summary of Observations and Data}

We select E+A galaxies from a major program to study four fields with
HST/WFPC2 imaging and ground-based spectroscopy.  The three lower
redshift fields are centered on X-ray luminous clusters at
$z=0.33,~0.58,~\&~0.83$, and the fourth on a QSO at $z=0.92$.  Our
dataset combines HST/WFPC2 mosaics of $6-12$ pointings in each field
with extensive redshift surveys completed primarily with Keck/LRIS
\citep{oke:95}.  From $\sim$1500 redshifts obtained in a total area of
$\sim200\Box'$, we isolate 797 field galaxies.  The observational
details of each field are in Table~\ref{fields}.  Here we describe
briefly the spectra and photometry used in this paper.

\subsection{Spectroscopy}

As a significant benefit of our large program to study galaxy
evolution in clusters, we have obtained redshifts for a large sample
of field galaxies to $z\sim2$.  During multiple observing runs from
June 1996 to July 2001, multi-slit spectra in the MS2053, MS1054, and
3C336 fields were obtained with Keck/LRIS.  Targets in the MS2053
field were selected using a magnitude cut of $m_{814}\leq23$, and in
both the MS1054 and 3C336 fields the magnitude cut was
$m_{814}\leq23.5$.  Spectra in the CL1358 field were collected at the
WHT and MMT where the primary targets were selected to have $R\leq21$.
Details of the target selection, spectral reduction, wavelength
calibration, sky subtraction, and removal of telluric absorption are
in \citet{fisher:98}, \citet{tran:99}, \citet{vandokkum:00},
\citet{tran:02}, and \citet{magee:04}.

\begin{figure}
\plotone{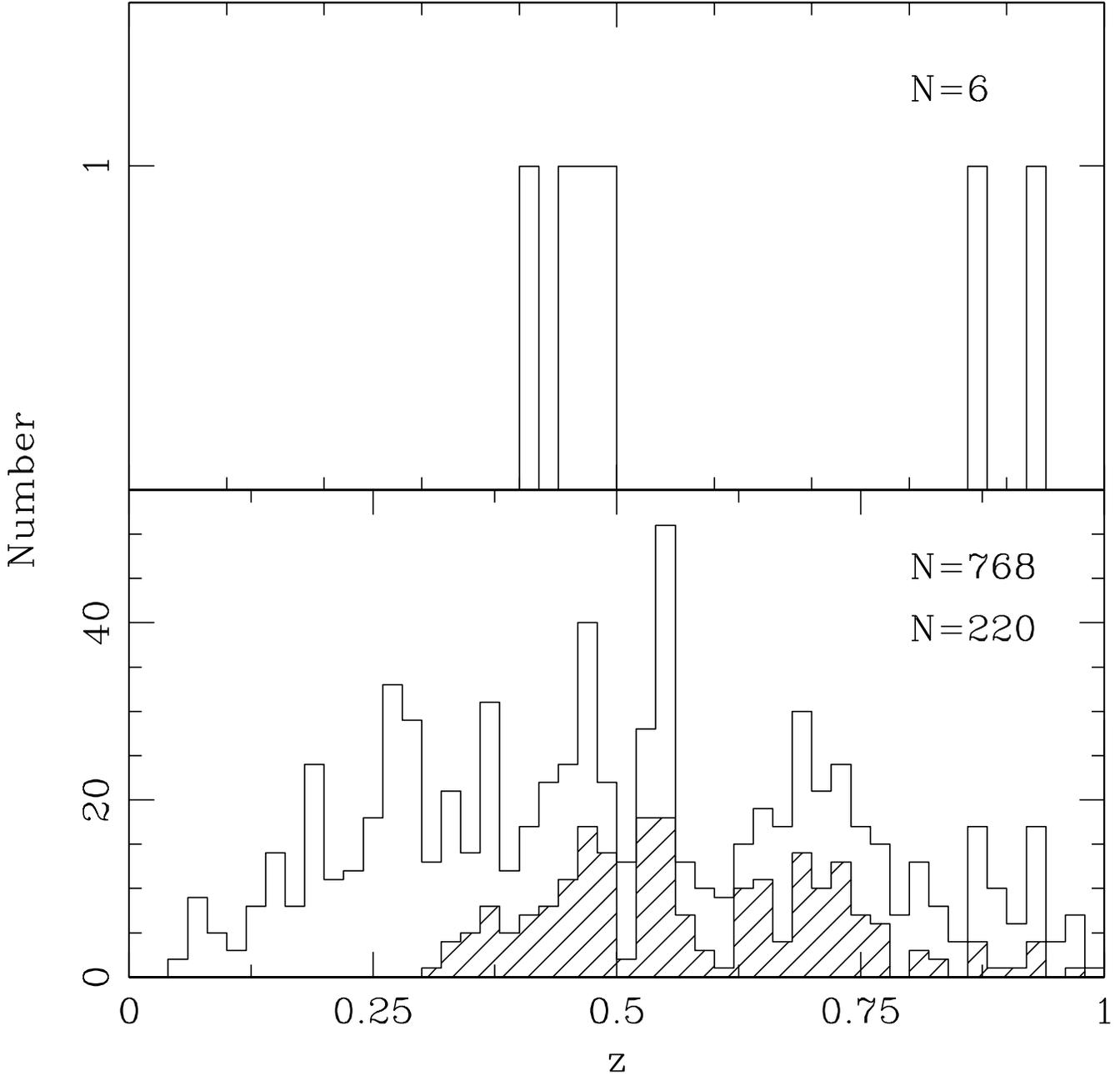}
\caption{ The redshift distribution of the six field E+A's (top) and
  that of all field galaxies at $0<z<1$ (768; bottom).  To determine
  the E+A fraction, we consider {\it only} field galaxies at $0.3<z<1$
  with $[S/N]_{BI}\geq20$ that have measurable [OII], H$\delta$, and
  H$\gamma$ (shaded region; 220 galaxies).  The field E+A fraction in
  this case is $2.7\pm1.1$\%; the field E+A redshift distribution is
  indistinguishable from that of the field sample using the K-S test.
\label{zhistall}}
\end{figure}

Galaxies were assigned redshifts using IRAF cross-correlation routines
as well as by visual inspection of each individual spectrum.  Each
redshift was given a quality flag where $Q=3, 2, \&~1$ corresponded to
definite, probable, and maybe (single emission line).  In our analysis,
we consider only galaxies with a redshift quality flag of $Q=3$; this
brings the cumulative field sample to 797 galaxies at $0.05<z<2.06$.
Figure~\ref{zhistall} (bottom panel) shows the redshift distribution of
our field sample for galaxies at $0<z<1$.  Note we do not include here
the $\sim500$ cluster members identified in these fields; all galaxies
with redshifts within $5\sigma$ of a given cluster's mean redshift were
excised from the field sample.

\subsection{HST/WFPC2 Imaging}

The four fields were imaged by HST/WFPC2 in the F606W and F814W
filters; each image mosaic was made of $6-12$ overlapping pointings.
The image reduction and photometry are detailed for CL1358, MS2053,
MS1054, and 3C336 in \citet{vandokkum:98a}, \citet{hoekstra:00},
\citet{vandokkum:00}, and \citet{magee:04} respectively.  Following
the method outlined in \citet{vandokkum:96}, we transform from the
WFPC2 filter system to redshifted Johnson magnitudes using:

\begin{equation}
\begin{array}{l}
B_z = F814W+a(F606W-F814W)+b \\
V_z = F814W+c(F606W-F814W)+d
\end{array}
\end{equation}

\noindent where the constants $\{a,b,c,d\}$ depend on the galaxy's
redshift and were calculated for an E/S0 galaxy spectral energy
distribution \citep{pence:76}.  $(B-V)_z$ colors were measured using a
$3''$ diameter aperture.  

\section{Field E+A Sample}

The E+A fraction depends strongly on how E+A galaxies are selected.
Although authors generally include [OII]$\lambda3727$ in their
selection criteria, they differ on which Balmer lines to use and
equivalent width limits to adopt
\citep[Z96]{dressler:83,couch:87,balogh:99,dressler:99}.  For example,
the E+A fraction in a given sample can differ by as much as a factor of
two if [OII] and only H$\delta$ are used as selection criteria compared
to a combination of [OII] and all three Balmer lines, and the E+A
fraction from requiring a Balmer equivalent width of $\leq-5$\AA~will
be smaller than that measured using $\leq-4$\AA.  Here we describe our
stringent E+A selection criteria and estimate the field E+A fraction at
intermediate redshifts ($0.3<z<1$).

\subsection{E+A Selection}

To determine the field E+A fraction at intermediate redshifts, we
define an average Balmer index

\begin{equation}
BI=(H\delta+H\gamma)/2.
\end{equation}

\noindent We consider only galaxies at $0.3<z<1$ with
$[S/N]_{BI}\geq20$, and for which [OII]$\lambda3727$, H$\delta$, and
H$\gamma$ were included in the spectral range.  These strict criteria
reduce the original sample of 797 galaxies to 220 (Fig.~\ref{zhistall},
bottom).  We select E+A galaxies as having
$($H$\delta+$H$\gamma)/2\leq-4$~\AA~and no significant [OII] emission
($<5$~\AA).  The bandpasses used to determine the equivalent widths of
these features are the same as those used in \citet{fisher:98} and
T03a.

Although combining the three Balmer lines to select E+A's is the most
robust approach \citep{newberry:90}, this is not possible for all the
galaxies due to the large redshift range covered by our survey.  In
addition, H$\beta$ becomes severely compromised by skylines at
$z\gtrsim0.8$.  As in our cluster survey (T03a), we apply a signal to
noise cut ($[S/N]_{BI}\geq20$) to the entire field sample to ensure the
E+A sample is not contaminated by nearby spectral types; this is
comparable to selecting galaxies with [OII], H$\delta$, and H$\gamma$
equivalent width errors of $\lesssim2$\AA.

\begin{figure}
\plotone{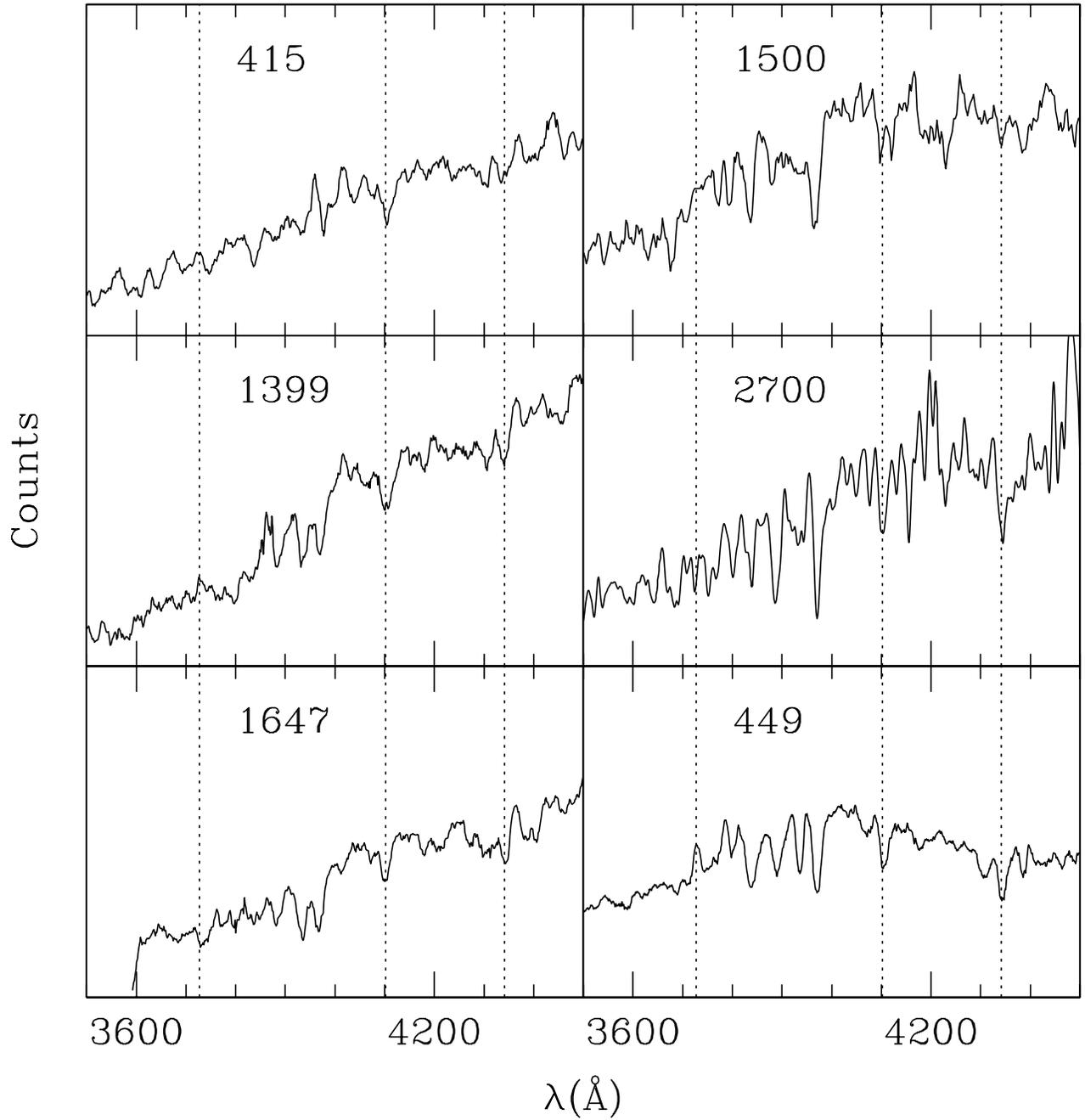}
\caption{Smoothed rest-frame spectra in arbitrary flux units of the
  six field E+A galaxies that satisfy our selection criteria of
  [OII]$\lambda3727<5$\AA, $($H$\delta+$H$\gamma)/2\leq-4$\AA, and
  $[S/N]_{BI}\geq20$.  The vertical dotted lines show [OII], H$\delta$,
  and H$\gamma$.
\label{spectra}}
\end{figure}

\begin{figure}
\epsscale{0.75}
\plotone{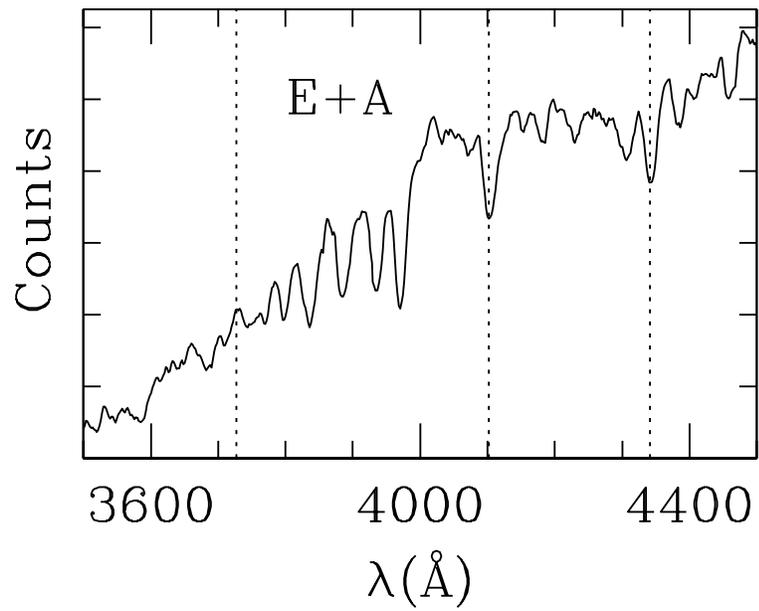}
\caption{Co-added rest-frame spectrum of the six field E+A's
  ($\bar{z}=0.6$) where the strong absorption in both H$\delta$ and
  H$\gamma$ is evident.  The vertical dotted lines are as in
  Fig.~\ref{spectra}.
\label{sum}}
\end{figure}

Using these selection criteria, we identify ten E+A candidates in our
field sample.  Visual inspection of these candidates find four to be
spurious due to, e.g., problems with removal of night sky features.  We
list the six field E+A galaxies that satisfy our stringent selection
process with their [OII] and Balmer equivalent widths in
Table~\ref{eqw}.  Their 1D spectra are shown in Fig.~\ref{spectra}, and
Fig.~\ref{sum} shows their co-added spectrum where the mean redshift is
$\bar{z}=0.6$.

\subsection{Field E+A Fraction}

The field E+A fraction at $0.3<z<1$ is $2.7\pm1.1$\%, a factor of four
lower than typically found in galaxy clusters at comparable
redshifts\footnote{Note that while the difference in the E+A fraction
between field and cluster environment is secure, the overall frequency
of E+A galaxies may vary depending on the luminosity limit of a given
sample. } ($11\pm3$\%; T03a).  We emphasize that because the field and
cluster E+A's were selected using the exact same criteria, we
circumvent problems associated with comparing between different
surveys.  The large difference in the field and cluster E+A fractions
at $0.3<z<1$ is therefore unlikely to be the result of selection
effects.

If we apply Z96's more conservative definition of an E+A galaxy
([OII]$<2.5$\AA, [H$\beta+$H$\gamma+$H$\delta]/3\leq-5.5$\AA) to our
sample, only one galaxy (2053-2700) would be considered an E+A.  In
this case, the post-starburst fraction of $\sim0.5$\% is comparable to
the low value measured by Z96 from the LCRS (0.2\%).  Such a low
fraction emphasizes the rarity of strong post-starburst galaxies even
in the intermediate redshift field.

One unexpected result is how strongly the E+A fraction varies between
the four fields.  Given four of the six E+A's lie in only one field
(MS2053), it is not so surprising then that different groups sampling
relatively small areas find widely varying E+A fractions for the
intermediate redshift field \citep[e.g.  $1-5$\%;][]{hammer:97,
balogh:99,dressler:99}.  This highlights the fact that large areas are
needed to study E+A galaxies because they are so rare.

\section{Physical Properties}

\begin{figure}
\epsscale{0.75}
\plotone{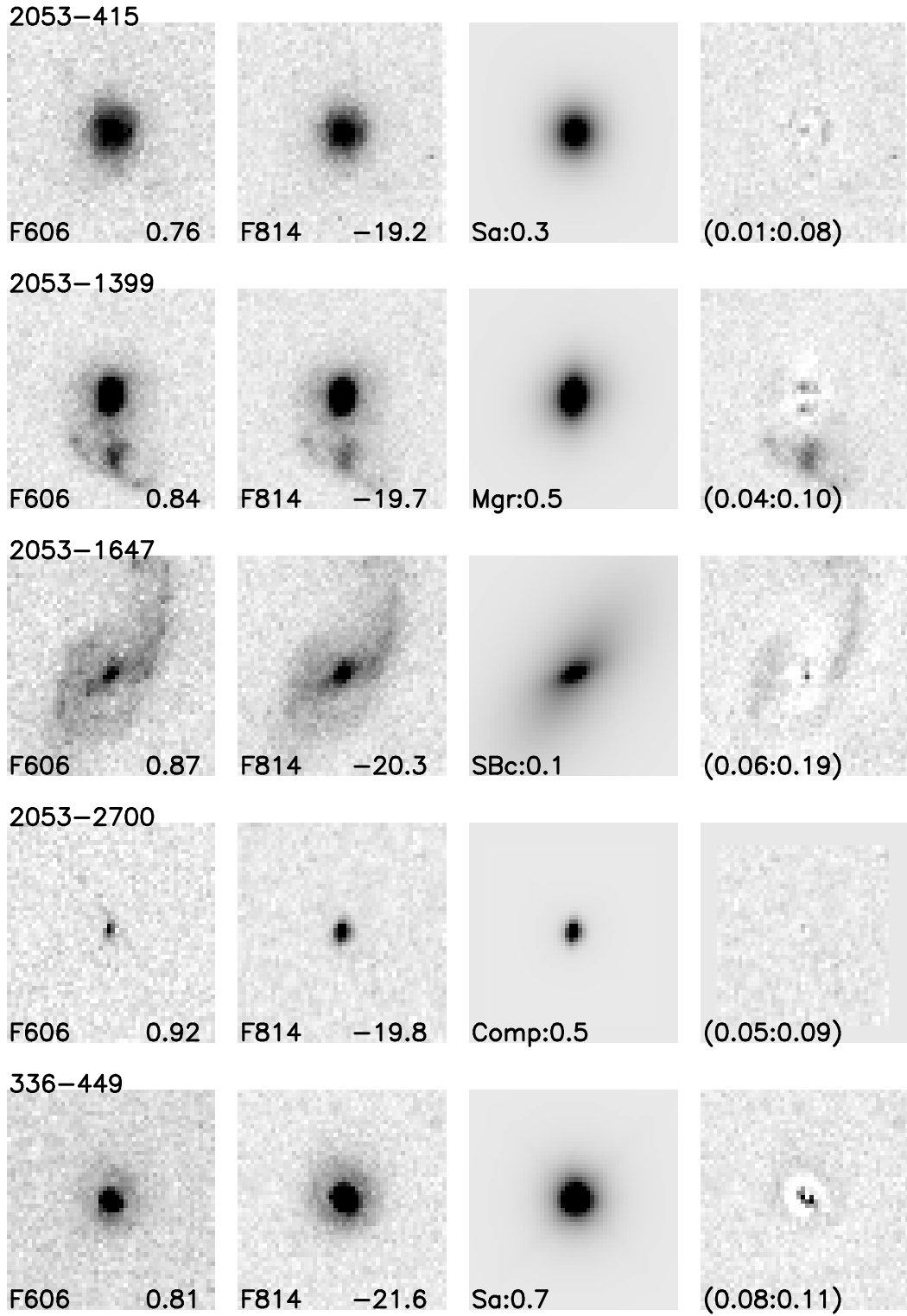}
\caption{Thumbnail images of the five field E+A's that fall on the
  HST/WFPC2 mosaics; the images are $5''\times5''$ and the E+A's
  ordered by increasing redshift.  For each E+A, we show (from left to
  right) the F606 image, F814 image, best-fit de Vaucouleurs
  bulge+exponential disk model in F814, and residual image in F814.
  Listed for each E+A is its $(B-V)_z$ color, absolute $B_z$
  magnitude, morphological type, bulge-to-total ratio ($B/T$), and
  galaxy residuals ($R_A:R_T$).  Three of the five E+A's (60\%) are
  bulge-dominated ($B/T\geq0.4$) systems, and the majority (80\%) are
  morphologically irregular ($R_A\geq0.05$ and/or $R_T\geq0.1$).
\label{gals}}
\end{figure}

By pairing HST/WFPC imaging with extensive spectroscopic surveys, we
have the unprecedented opportunity to study in detail the physical
properties of the field E+A population at $z>0.3$.  The rarity of these
objects has been a formidable obstacle in studying this population, so
it is no surprise that the physical characteristics of intermediate
redshift field E+A's are virtually unknown.  Here we examine their
luminosities, colors, structural parameters, estimated internal
velocity dispersions, and environment to better understand the origin
of the E+A phase and its role in galaxy evolution.  The properties of
the five E+A's that fall on the HST/WFPC2 imaging are listed in
Table~\ref{properties}, and their thumbnail images are shown in
Fig.~\ref{gals}.

\subsection{Luminosity}

Previous studies of E+A galaxies in both the field and in clusters
suggest that E+A's must have a heterogenous parent population because
of their wide range in luminosity
\citep[e.g.][Z96]{wirth:94,caldwell:99}.  We cannot address this issue
as our spectroscopic limits imply that we are biased towards the
brightest E+A galaxies.  However, we note that the E+A's in our sample
include very luminous systems ($-21.6\leq M_{Bz}-5\log h\leq -19.2$);
even if the brightest E+A's fade by as much as $\sim1.5$ mag
\citep{couch:87,barger:96}, they will remain $L>L^{\ast}$ systems.

\subsection{Color} 

\begin{figure}
\plotone{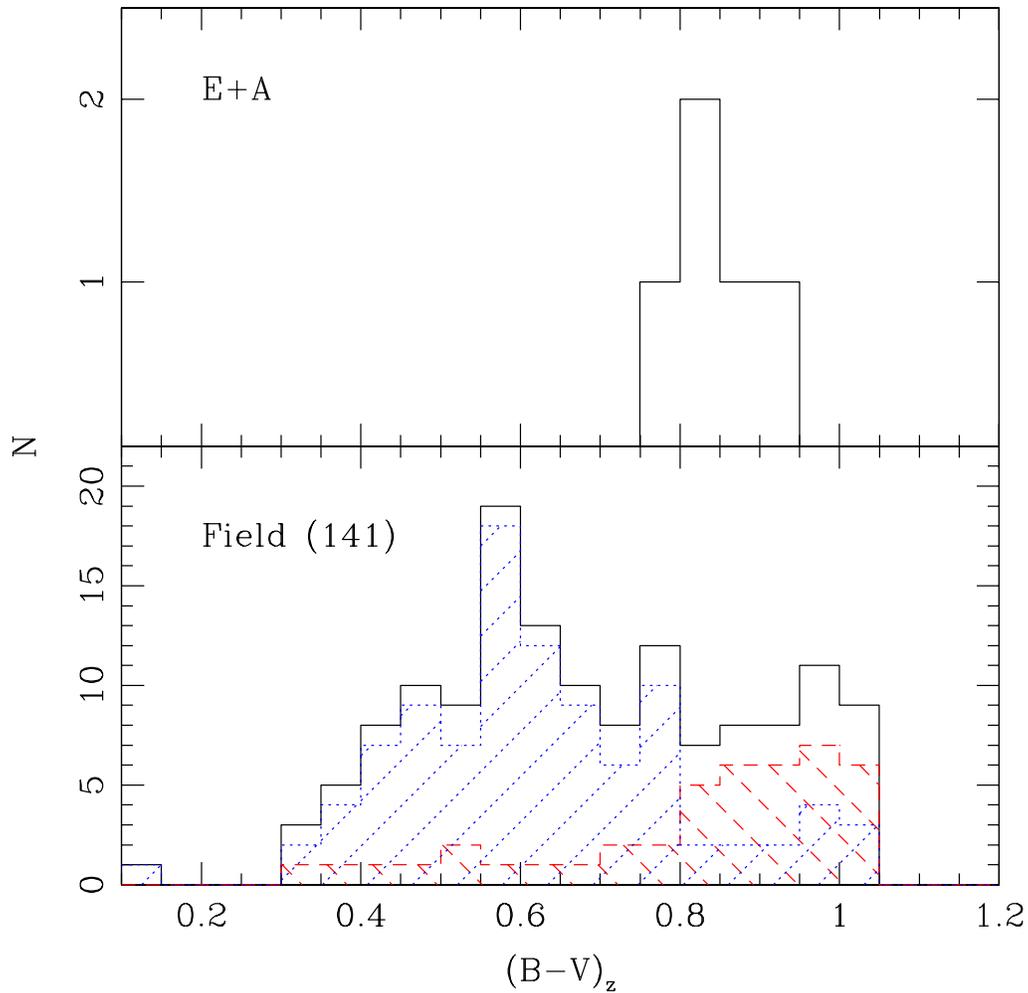}
\caption{$(B-V)_z$ distribution of the five field E+A's (top) and 141
  field galaxies at $0.3<z<1$ (bottom, solid line; see \S3.1 for
  selection) that fall on the HST/WFPC2 mosaics; colors are determined
  from imaging in F606 and F814.  The field sample is split into
  absorption ([OII]$<5$\AA; dashed-hatched regions) and emission
  ([OII]$\geq5$\AA; dotted-hatched regions) line galaxies.  The E+A's
  are redder than the average field galaxy; the field E+A color
  distribution differs from that of the field sample with $>95$\%
  confidence (K-S test).  Comparison of the E+A's to the emission line
  galaxies finds their color distributions to differ with 99.7\%
  confidence, but there is no difference between the E+A's and the
  absorption line galaxies.  These results are consistent with the
  E+A's evolving into the redder absorption line systems.
\label{BVhist}}
\end{figure}

From the HST/WFPC2 imaging, we measure $(B-V)_z$ for 141 field
galaxies (of 220; see \S3.1 for selection) and five E+A's; the color
distribution of both populations are shown in Fig.~\ref{BVhist}.
There is a striking difference in the E+A color distribution compared
to the field: E+A's are uniformly {\it redder} than the blue,
star-forming galaxies that make up the majority of the field.  Using
the Kolmolgorov-Smirnov test \citep{press:92}, we find the E+A color
distribution differs from that of the field sample with $>95$\%
confidence.  If we consider only the emission line galaxies
([OII]$\geq5$\AA), the E+A color distribution differs with 99.7\%
confidence, whereas the E+A color distribution is indistinguishable
from that of the absorption line galaxies ([OII]$<5$\AA).

One might expect the E+A's to be blue since the E+A phase is associated
with a recent starburst.  However, models show that E+A's remain as
blue as actively star-forming galaxies for only a fraction of the total
time the E+A phase is visible \citep[$\sim0.5$ Gyr vs. 1.5
Gyr;][]{barger:96}.  Thus E+A's can evolve quickly away in color from
the blue envelope of field galaxies at $(B-V)_z\sim0.6$ to join the red
population at $(B-V)_z\sim0.9$.  The colors of our E+A galaxies are
consistent with them being the progenitors of some absorption line
galaxies.  The fact that these E+A's are red indicates that colors
alone are not sufficient to identify galaxies that were recently star
forming systems.

\subsection{Half-light Radius}

For galaxies that fall on the HST/WFPC2 imaging, our analysis includes
structural parameters determined from 2D de Vaucouleurs
bulge+exponential disk decompositions fit with GIM2D
\citep{simard:99,simard:02}.  From these surface brightness fits, we
obtain bulge-to-total ratios ($B/T$) and half-light radii, as well as
the galaxy residuals $R_A$ and $R_T$.  $R_A$ and $R_T$ are the
residuals contained in an asymmetric and symmetric component,
respectively, and quantify how much a galaxy's profile deviates from a
bulge+disk model; they are measured by subtracting the best-fit model
from the galaxy image \citep{tran:01,tran:03a}.

The field E+A's include both compact ($r_{1/2}=0.8$\hi kpc) systems and
large disks ($r_{1/2}\sim8$\hi kpc).  The half-light radii for four of
the E+A's is comparable to values measured for E+A's in clusters
($\sim1-4$\hi kpc; T03a).  However, we note the inclusion of one
unusually large galaxy (2053--1647; $r_{1/2}\sim8$\hi kpc) that is also
the most disky E+A in our sample.  Because our spectroscopy is biased
towards the central regions, the E+A phenomenon may be
localized to 2053--1647's ``bulge''; only with fully integrated spectra
can we determine if the entire galaxy is going through an E+A phase.

\subsection{Morphology}

From the de Vaucouleurs bulge+exponential disk fits, we find the E+A's
are more likely to be bulge-dominated systems
\citep[$(B/T)_{deV}\geq0.4$;][]{tran:01} than the average field galaxy:
60\% of the E+A's are bulge-dominated compared to $\sim30$\% of the
field \citep{simard:02}. Interestingly, both \citet{quintero:04} and
\citet{yang:04} also find that nearby post-starburst galaxies ($z<0.2$)
tend to be bulge-dominated systems.

All the E+A's have bright, centrally concentrated light profiles,
similar to what \citet{caldwell:99} find in their study of
post-starburst galaxies in Coma.  From the smooth light profiles of
their E+A's, \citet{caldwell:99} suggested that the starburst preceding
the E+A phase was probably an event localized in the galaxy's center
and not due to, e.g., clumpy star-forming regions spread throughout a
larger disk.  However, removal of the 2D model from the E+A's in our
field sample shows that three (2053-415, 2053-1399, \& 2053-1647) have
clumpy, non-axisymmetric structure in addition to a bright central
region (Fig.~\ref{gals}, right).  Unlike the E+A's in Coma, our field
E+A's show both centralized bright regions and extended, clumpy star
formation.  As noted in \S4.3, the E+A phenomenon may be associated
primarily with the ``bulge'' component in these systems.

\subsection{Evidence of Recent Interactions}

Past studies of E+A's in the nearby field suggest the E+A phenomenon is
associated with galaxy-galaxy encounters \citep[Z96]{liu:95,yang:04}.
If the E+A phase was triggered by a recent interaction, we would expect
the E+A's to be morphologically disturbed because such features remain
visible for the duration of the E+A phase \citep{mihos:94a,mihos:94b}.
We find four of the five field E+A's with WFPC2 imaging have elevated
degrees of galaxy asymmetry ($R_A\geq0.05$) and/or total residuals
\citep[$R_T\geq0.1$;][]{schade:96,tran:01}; one is even visually typed
as a merger.  These results are consistent with the E+A phase being
associated with galaxy-galaxy interactions.

\subsection{Estimated Internal Velocity Dispersions}

Because internal velocity dispersion ($\sigma$) is a useful tracer of
the galaxy's total mass, we estimate $\sigma_{est}$ for the E+A's that
have $(B-V)_z$ and structural parameters measured from HST/WFPC2
imaging.  As described in \citet{kelson:00c} and T03a, we can use
colors to correct the mass-to-light ratios of the galaxies and so
estimate the internal velocity dispersions of these systems.  In this
method, we essentially evolve the E+A's by fading and reddening them
until they lie on the color magnitude relation (CMR) defined by a
passively evolving galaxy population; here we use the CMR normalized to
the early-type galaxies in MS1054 ($z=0.83$).  We also correct the
E+A's for simple fading as determined from the Fundamamental Plane
\citep[$\Delta\log (M/L)\propto -0.40z$;][]{vandokkum:98a}.

We find the field E+A's span a large range in estimated internal
velocity dispersion: $50\lesssim\sigma_{est}\lesssim220$\kms~
(Table~\ref{properties}).  The two field E+A's with the largest
$\sigma_{est}$ are also the two highest redshift systems.  It may be
that 1) the mass distribution of field E+A's evolves with redshift
and/or 2) the fraction of field E+A's increases with redshift.  It is
interesting to note that there are at least a few field E+A's at
$z<0.13$ with measured $\sigma>200$\kms~\citep{norton:01}.  However,
the current sample sizes are too small to determine whether the field
E+A dispersion distributions differ between low and intermediate
redshifts.

\subsection{Environment}

\begin{figure}
\plotone{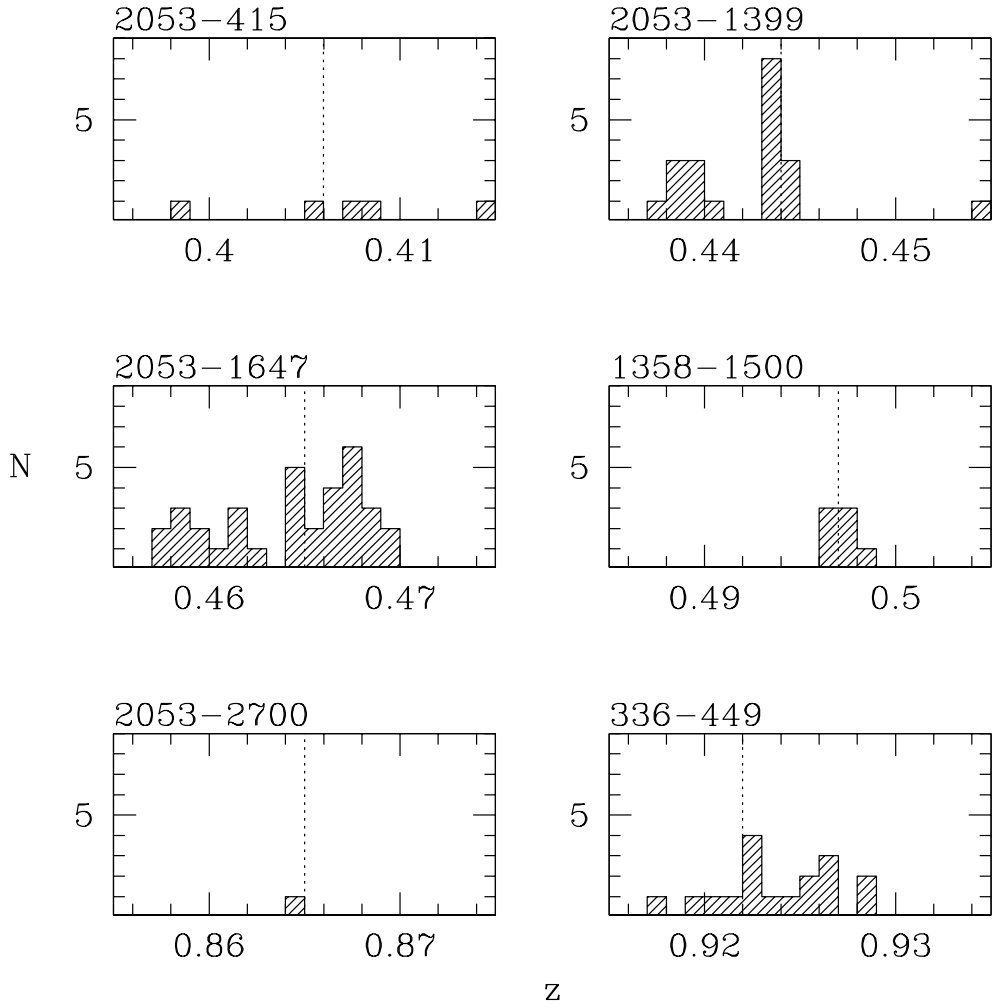}
\caption{ Utilizing our entire field sample,  we show the redshift
  distributions of galaxies in the same fields as the E+A's; the
  vertical dotted line denotes the E+A's redshift.  The average field
  E+A has $6.3\pm1.0$ redshift neighbors ($(cz)_{rest}\leq500$\kms), a
  number comparable to that of an average field galaxy in our survey
  ($4.9\pm0.1$).  This indicates that like most field galaxies, E+A's 
  tend to lie in galaxy groups.
\label{zhist_ea}}
\end{figure}

Z96 originally suggested that most E+A's lie in galaxy groups because
galaxy-galaxy interactions occur more frequently in groups than in the
field.  Because the majority of field galaxies are in groups, we would 
expect that most
of the E+A's are also in groups.  To test whether our E+A's lie in
environments similar to that of a typical field galaxy, we use our
entire field redshift sample at $0.3<z<1$ to estimate the average
number of redshift neighbors each E+A has.  Here we consider a galaxy
to be a neighbor if its redshift is within $(cz)_{rest}\leq500$\kms~of
the E+A and it is in the same field; the redshift distribution of
galaxies near each E+A is shown in Fig.~\ref{zhist_ea}.  The average
field E+A in our survey has $6.3\pm1.0$ neighbors, a value consistent
with the number of neighbors an average field galaxy has ($4.9\pm0.1$);
here we asssume a Poissonian distribution to estimate the errors.  This
indicates that like most field galaxies, field E+A's tend to 
lie in galaxy groups even at intermediate redshifts.


\section{Discussion}

Because the E+A galaxies are found in both the field and clusters at
redshifts up to $z\sim1$, they can place constraints on galaxy
formation models.  Models must not only produce E+A's but also how the
E+A fraction varies with environment.  Here we discuss in greater
detail the importance of E+A galaxies and how they provide interesting
constraints for galaxy evolution models.

\subsection{Where Are the E+A's?}

Like E+A galaxies in the nearby universe \citep[Z96;][]{quintero:04},
the majority of E+A's at $z>0.3$ are in the field/group environment.
Assuming 10\% of galaxies are in clusters \citep{gomez:03} and the E+A
fraction in clusters and the field is 10\% and 3\% respectively, we
find $(N_{field}/N_{cluster})\sim3$.  However, the elevated E+A {\it
  fraction} in clusters indicates environment does have an important
role in triggering the E+A phase.  Most likely additional cluster
processes, e.g. galaxy harassment and/or interactions with a hot
intracluster medium \citep{caldwell:99,poggianti:04}, can also trigger
an E+A phase.

\subsection{Importance of the E+A Phase}

To determine whether the E+A phase has an significant role in the
conversion of emission line galaxies into absorption line systems, we
estimate the fraction of absorption line systems in the field that
have undergone an E+A phase by $z=0$.  In our field sample of 220
galaxies at $0.3<z<1$, we have 66 absorption line ([OII]$<5$\AA) and
154 emission line ([OII]$\geq5$\AA) systems.  The field E+A fraction
when considering only absorption line galaxies is $\sim9$\%.
Combining this fraction with the elapsed time between $z=0.3$ and
$z=1$ ($\Delta t=4.3$ Gyr; $H_0=70$~\kms~Mpc$^{-1}$, $\Omega_M=0.3$,
$\Lambda=0.7$) and assuming the E+A phase is visible for 1.5 Gyr
\citep{couch:87,barger:96}, we estimate $\sim25$\% of absorption line
galaxies in the local field had an E+A phase at $0<z<1$.  These
results show that E+A's can be an important link in the evolution of
star-forming galaxies into passive systems in the field.

If all E+A's evolve into early-type galaxies, they play a vital role
in the evolution of these objects.  If we consider only the field
galaxies morphologically typed as early-types (E/S0; 24), the fraction
that have undergone an E+A phase since $z\sim1$ increases to
$\sim70$\%.  Thus E+A's can be as important to the evolution of
early-type galaxies in the field as in galaxy clusters.

It is interesting to compare the fraction of early-type galaxies that
have undergone an E+A phase at $z<1$ to early-type field galaxies at
$z\sim1$.  From a co-added spectrum of 10 early-types in the HDF-N
($\bar{z}=0.9$), \citet{vandokkum:03b} find enhanced H$\delta$
absorption and postulate that these early-types experienced a recent
starburst.  Our estimate that $\sim25-70$\% of early-type galaxies in
the nearby field had an E+A phase since $z<1$ shows that early-types
can have starbursts at $z<1$ as well.

\subsection{Major versus Minor Mergers}

The high fraction of morphologically irregular field E+A's ($\sim80$\%)
suggests these systems are associated with galaxy-galaxy interactions,
e.g minor or major mergers.  Here we attempt to distinguish between the
two types of interactions.  All the field E+A's have the bright,
centrally concentrated light profiles normally associated with minor
mergers \citep{mihos:94a} but some also have the clumpy, extended star
forming regions that are usually created in major mergers
\citep{barnes:96}.  Using their bulge/disk values, it may be possible
to discriminate between these two viable mechanisms: minor mergers can
trigger a central starburst but allow the disk to survive
\citep{walker:96} whereas major mergers effectively destroy any disk
component \citep{barnes:92,barnes:96,naab:99}.

Of the five E+A's in our sample with WFPC2 imaging, four have
significant disk components ($B/T\lesssim0.5$).  The retention of
these disks suggests that minor mergers are the primary driver of the
E+A phase in the field.  However, we stress that a considerably larger
field E+A sample is needed to test this hypothesis since, e.g., we may
not be observing the E+A's in their final morphological state.

\subsection{E+A Frequency:  A Constraint on Formation Models}

We find the field E+A fraction at $0.3<z<1$ is $\sim3$\%, but is this
too low for current galaxy formation models?  Hierarchical models
conservatively predict $\sim20$\% of field galaxies ($L\gtrsim
L^{\ast}$) have had a minor/major merger since $z\sim1$ \citep[$\Delta
t_{z=1\rightarrow0}\sim8$ Gyr]{lacey:94,somerville:99a}.  Assuming 1)
5\% of field galaxies had a galaxy-galaxy interaction at $0.3<z<1$
($\Delta t=4.3$ Gyr) within the last 2 Gyr of it being observed and 2)
that these interactions trigger strong episodes of star formation
\citep{mihos:94a,mihos:94b}, we would expect to have $\sim11$
starburst/post-starburst galaxies in our sample.  Using a timescale of
0.5 Gyr for the starburst and 1.5 Gyr for the post-starburst phase
\citep{dressler:83,barger:96} implies we should have $\sim8$ field
E+A's, a value consistent with the number we actually find.

While our approach has several caveats, this exercise illustrates how
the field E+A fraction at intermediate redshifts can be an interesting
constraint on galaxy formation models.  A viable model cannot
significantly over/under-predict the field E+A fraction, and it must
account for the $\gtrsim25$\% of absorption line galaxies in the
nearby field that had an E+A phase at $z<1$.  It should also account
for the $\gtrsim30$\% of early-type galaxies in clusters that had an
E+A phase at $z<1$ (T03a).

\section{Conclusions}

We select E+A galaxies from a spectroscopic dataset of $\sim800$ field
galaxies and measure the E+A fraction in the intermediate redshift
field ($0.3<z<1$) to be $2.7\pm1.1$\%.  Because we select the field
E+A's in the exact same manner as E+A's in galaxy clusters at
comparable redshifts, we can compare directly how the E+A fractions
differ between the two environments.  We find the E+A fraction in the
field is a factor of four smaller than that observed in intermediate
redshift galaxy clusters ($11\pm3$\%; T03a).  However, most E+A
galaxies ($\sim70$\%) lie in the field because only a small fraction
of galaxies are in clusters.

From HST/WFPC2 imaging, we present first results on the physical
properties of field E+A's at $z>0.3$.  The five E+A galaxies that fall
on the WFPC2 mosaics include luminous systems ($-21.6\leq M_{Bz}-5\log
h\leq -19.2$), and they span a range half-light radius ($0.8\lesssim
r_{1/2}<8.2$\hi kpc) and estimated internal velocity dispersion
($50\lesssim\sigma_{est}\lesssim220$\kms).  These properties indicate
that like their lower redshift counterparts, the E+A's in the
intermediate redshift field have a heterogeneous parent population.

Despite being descended from a variety of progenitors, these E+A's have
several characteristics in common.  Compared to the general field
population, the E+A's are uniformly redder than the blue, star-forming
galaxies that make up the majority of the field.  These E+A's also are
more likely than the average field galaxy to be a bulge-dominated
system, and they tend to be morphologically irregular.  These
characteristics indicate the E+A phase is associated with galaxy-galaxy
interactions, and that E+A's are an important link between star-forming
and passive galaxies.

We find E+A's make up $\sim9$\% of the absorption line galaxies at
$0.3<z<1$.  Combining this with timescale arguments, we estimate that
$\sim25$\% of passive galaxies in the nearby field had an E+A phase at
$z<1$.  This fraction increases to $\sim70$\% if we consider only
galaxies that are morphologically typed as E/S0's.  Our study
demonstrates that the E+A phase can have as significant a role in how
galaxies evolve into early-type systems in the field as in galaxy
clusters.  However, we recognize our results are based on a small
sample of E+A's, and we stress that additional studies of this
interesting population are needed.

\acknowledgments

This research has been supported in part by the Swiss National Science
Foundation.  Additional support from NASA HST grants GO-06745.01,
GO-07372.01, and GO-08220.03 are gratefully acknowledged.  P.  van
Dokkum also acknowledges support from STScI grant HST-AR-09541.01-A.
The authors thank the entire staff of the W. M.  Keck Observatory for
their support, and extend special thanks to those of Hawaiian ancestry
on whose sacred mountain we are privileged to be guests.

\bibliographystyle{/home/vy/aastex/apj}
\bibliography{/home/vy/aastex/tran.bib}

\clearpage

\clearpage

\begin{deluxetable}{lrrrrrl}
\tablecolumns{6}
\tablewidth{0pc}
\tablecaption{Observational Details\label{fields}}
\tablehead{
\colhead{Field} & \colhead{($\alpha,~\delta)_{2000}$}& 
\colhead{$N_{F}$~\tablenotemark{a}} &
\colhead{$N'_{F}$~\tablenotemark{b}} &
\colhead{$N_{E+A}$} & \colhead{\%$_{E+A}$} & \colhead{Reference}}
\startdata
CL~1358+62  & (13:59:50.7, 62:31:05.4) & 156 & 76 & 1 & 1.3&\citet{fisher:98}\\
MS~2053--04 & (20:56:21.4, --4:37:51.5)& 215 & 83 & 4 & 4.8& \citet{tran:02}\\
MS~1054--03 & (10:57:00.0, --3:37:37.0)& 186 & 42 & 0 & 0.0& \citet{tran:99}\\
            &                          &     &  & & &\citet{vandokkum:00}\\
            &                          &     &  & & &\citet{tran:02}\\
3C336       & (16:24:39.1, 23:45:12.1) & 240 & 19 & 1 & 5.3&\citet{magee:04}\\
            &                          & 797 &220 & 6 & 2.7& $Total$\\
\enddata
\tablenotetext{a}{The total number of spectroscopically confirmed field
galaxies.  We do {\it not} include cluster galaxies in this number for
the three cluster fields.} 
\tablenotetext{b}{The number of field galaxies in the redshift range
  $0.3<z<1$ with spectral $[S/N]_{BI}\geq20$, and whose spectra include
  [OII]$\lambda3727$, H$\delta$, and H$\gamma$.} 
\end{deluxetable}

\begin{deluxetable}{lrrrrr}
\tablecolumns{6}
\tablewidth{0pc}
\tablecaption{Redshifts and Equivalent Widths\tablenotemark{a}\label{eqw}}
\tablehead{
\colhead{Galaxy} &  \colhead{$z$} & \colhead{[OII]} &
\colhead{H$\delta$} & \colhead{H$\gamma$} & \colhead{H$\beta$} }
\startdata
2053--415 &0.406 & $4.5\pm2.1$ & $-6.0\pm1.5$ & $-4.1\pm1.3$ & $-0.4\pm1.1$ \\
2053--1399 &0.444 & $2.7\pm1.1$ & $-6.1\pm0.7$ & $-2.6\pm0.7$ & $-4.2\pm0.5$ \\
2053--1647 &0.465 & $-2.4\pm2.1$ & $-4.2\pm1.5$ & $-4.8\pm1.4$ & $0.3\pm0.9$ \\
1358--1500 &0.497 & $1.5\pm1.8$ & $-4.7\pm2.2$ & $-4.1\pm1.9$ & $0.5\pm5.6$ \\
2053--2700 &0.865 & $0.0\pm2.3$ & $-3.4\pm1.8$ & $-7.8\pm1.9$ & $-8.3\pm2.1$ \\
336--449 &0.922 & $3.2\pm0.8$ & $-3.8\pm1.3$ & $-6.5\pm1.4$ & $-7.5\pm2.7$ \\
\enddata
\tablenotetext{a}{Bandpasses used to measure equivalent widths are the same as
those used in \citet{tran:03b} and \citet{fisher:98}.  Here positive values
denote emission and negative values absorption.}
\end{deluxetable}

\begin{deluxetable}{lrrrrrrrrrr}
\tablecolumns{11}
\tablewidth{0pc}
\tablecaption{Physical Properties\tablenotemark{a}\label{properties}}
\tablehead{
\colhead{Galaxy} &  \colhead{$(m-M)$\tablenotemark{a}} & 
\colhead{$B_z$}  &  \colhead{$M_{B}$}   &\colhead{$(B-V)_z$} &
\colhead{T-type\tablenotemark{b}} &  \colhead{$(B/T)$\tablenotemark{c}} &
\colhead{$r_{1/2}$\tablenotemark{c}}    &\colhead{$R_A$\tablenotemark{c}} &
\colhead{$R_T$\tablenotemark{c}} &\colhead{$\sigma_{est}$\tablenotemark{d}}}
\startdata
2053--415 & 40.9 & 21.7 & --19.2 & 0.76$\pm0.01$ & Sa & 0.3 & 2.1 & 0.01 & 0.08 & 51\\
2053--1399 & 41.2 & 21.5 & --19.7 & 0.84$\pm0.01$ & Mgr & 0.5 & 1.8 & 0.04 & 0.10 & 121\\
2053--1647 & 41.3 & 21.0 & --20.3 & 0.87$\pm0.01$ & SBc & 0.1 & 8.2 & 0.06 & 0.19 & 68\\
1358--1500 & 41.5 & \nodata & \nodata & \nodata & \nodata & \nodata & \nodata & \nodata & \nodata & \nodata\\
2053--2700 & 42.9 & 23.1 & --19.8 & 0.92$\pm0.01$ & Compact & 0.5 & 0.8 & 0.05 & 0.09 & 210\\
336--449 & 43.1 & 21.5 & --21.6 & 0.81$\pm0.00$ & Sa & 0.7 & 1.4 & 0.08 & 0.11 & 219\\
\enddata
\tablenotetext{a}{Distance moduli determined for $H_0=100$\kms Mpc$^{-1}$
$\Omega_M=0.3$, and $\Lambda=0.7$ cosmology.}
\tablenotetext{b}{Morphologically typed by \citet{fabricant:04} and
  K. Tran.}
\tablenotetext{c}{Structural parameters determined by fitting a de
  Vaucouleurs bulge+exponential disk to the galaxy's surface brightness profile
  \citep{tran:02,tran:03a}.  Half-light radii are in kpc.}
\tablenotetext{d}{Internal velocity dispersions are in \kms~and are estimated 
using the method outlined in \citet{tran:03b}.}

\end{deluxetable}

\end{document}